\documentclass[aps,pre,groupedaddress,11pt,tightenlines,nofootinbib]{revtex4-1}
\usepackage[hmargin=1.5in]{geometry}
\usepackage[T1]{fontenc}

\pdfoutput=1
\usepackage[dvipsnames,rgb,dvips,hyperref]{xcolor}
\usepackage{graphicx}
\usepackage{float}
\usepackage{overpic}
\usepackage{amsmath,amssymb,amsfonts,bbm}
\usepackage{hyperref}
\definecolor{linkblue}{HTML}{2200CC}
\hypersetup{
    colorlinks,
    linkcolor=linkblue,
    citecolor=linkblue,
    urlcolor=linkblue
}

\makeatletter
\def\amsbb{\use@mathgroup \M@U \symAMSb}
\makeatother
\usepackage[bbgreekl]{mathbbol}
\usepackage{mathrsfs}
\usepackage[most]{tcolorbox}
\newtcolorbox{inlinebox}[1][]{blanker,after=, 
     left=5mm, right=0mm, top=1mm, bottom=2mm,
     borderline west={1pt}{3mm}{gray}, #1}

\newcommand{\fourier}[1]{\ensuremath{\mathcal F #1}}
\newcommand{\invfourier}[1]{\ensuremath{\mathcal F^{-1} #1}}

\newcommand\nn{\nonumber}
\newcommand\ve[1]{\boldsymbol{#1}}

\newcommand{\rd}{\ensuremath{\textrm{d}}}
\newcommand{\tr}{\ensuremath{^{\rm T}}}

\newcommand{\Wi}{\ensuremath{\textrm{Wi}}}
\newcommand{\Ca}{\ensuremath{\textrm{Ca}}}

\newcommand{\eqnlab}[1]{\label{eqn:#1}}

\newcommand{\eqnref}[1]{(\ref{eqn:#1})}

\newcommand{\Eqnref}[1]{Eq.~(\ref{eqn:#1})}

\newcommand{\seclab}[1]{\label{sec:#1}}

\newcommand{\Secref}[1]{Section~\ref{sec:#1}}

\begin{document}

\title{Computer Algebra for Microhydrodynamics}

\author{Jonas Einarsson}
\date{\today}

\begin{abstract}
I describe a method for computer algebra that helps with laborious calculations typically encountered in theoretical microhydrodynamics. The program mimics how humans calculate by matching patterns and making replacements according to the rules of algebra and calculus.
This note gives an overview and walks through an example, while the accompanying code repository contains the implementation details, a tutorial, and more examples. The code repository is attached as supplementary material to this note, and maintained at\\\url{https://github.com/jeinarsson/matte}
\end{abstract}

\maketitle

\section{Introduction}

Microhydrodynamics concerns the motions of fluids and particles at low Reynolds numbers, where viscous forces dominate over inertial. In this regime the flow velocity and pressure fields $\ve u$ and $p$ are governed by Stokes equations
\begin{align}
  \nabla^2\ve u - \nabla p = 0\,,\quad \nabla \cdot \ve u = 0\,.\eqnlab{stokes}
\end{align}
Although \Eqnref{stokes} is linear, solving it is often complicated by of one or more of
\begin{itemize}
  \item boundary conditions imposed on geometry in which \Eqnref{stokes} is not separable,
  \item non-linear boundary conditions on fluid-fluid (or more complicated) interfaces,
  \item non-linear forcing due to non-Newtonian, inertial, or other non-linear effects.
\end{itemize}
One option is to produce a numerical solution to the governing equations. When the nonlinearities dominate the solutions that may be the only option. Theoretical progress, however, may often be made by perturbation theory when there is a small parameter controlling the magnitude of the nonlinearity. 

Perturbation problems describe the first deviations from simple Stokes flow due to some nonlinearity. For example, the first effects of viscoelasticity is described by a {\lq}second-order fluid{\rq} constitutive equation which renders the fluid stress non-linear in the flow velocity gradients. Weak non-linear effects that directly affect the momentum balance in the fluid generally result in an equation of the form
\begin{align}
  \nabla^2\ve u - \nabla p = \epsilon \ve f(\ve u)\,,\quad \nabla \cdot \ve u = 0\,,\eqnlab{nonlinear}
\end{align}
where $\ve f$ is any functional and $\epsilon$ is a small parameter.

Another example is when considering fluid-fluid interfaces with surface tension, for example a drop suspended in fluid. In this case there are boundary conditions on the stress jump across the interface, applied at the unknown position of the interface:
\begin{align}
  \Ca(\ve \sigma - \ve{\hat \sigma})\cdot \ve n =  \ve n(\nabla\cdot\ve n)\,,\quad \textrm{on the interface,}
\end{align}
where $\ve \sigma$ and $\ve{\hat \sigma}$ are the stress tensors on the two sides of the interface, and $\ve n$ is the unknown normal vector of the drop. This perturbation is controlled by a Capillary number $\Ca$, which in the limit of $\Ca = 0$ decouples the interface shape from the flow equations.

In perturbation theory we seek solutions to these problems on the form of a series, for example
\begin{align}
  \ve u = \ve u^{(0)} + \epsilon \ve u^{(1)} + \epsilon^2 \ve u^{(2)} + ...\,,\quad p = p^{(0)} + \epsilon p^{(1)} + \epsilon^2 p^{(2)} + ...\,,
\end{align}
and similarly for all relevant variables%
\footnote{%
Many perturbation problems in fluid mechanics are singular, and typically require multiple expansions and asymptotic matching. In those cases the examples in this note typically correspond to a {\lq}near field{\rq} problem in the vicinity of a boundary.
}%
. By inserting the series into the governing equations and comparing order by order in $\epsilon$ we typically find a sequence of equations that begins with a simple Stokes flow problem, yielding $\ve u^{(0)}$, $p^{(0)}$, and any other lowest order variables. Then follows a sequence of linear, but inhomogenous equations and/or boundary conditions for the higher orders, for example
\begin{align}
  \nabla^2\ve u^{(n)} - \nabla p^{(n)} = \ve f(\ve u^{(n-1)})\,,\quad \nabla \cdot \ve u^{(n)} = 0\,,\quad n>0\,.\eqnlab{perturbation_example}
\end{align}
Even for low values of $n$, evaluating $\ve f(\ve u^{(n-1)})$ can be tedious. Solving for $\ve u^{(n)}$ is even worse, and even if one avoids that by use of a reciprocal theorem the resulting integrals require substantial mathematical labor.
Although this labor is straightforward in principle, the sheer amount of work may be discouraging, and prone to error if carried through. Nevertheless, in order to test our models - the perturbation equations and their associated assumptions about physical reality - we must produce a prediction.

In this note and the accompanying code I describe a set of software routines, implemented in Mathematica, that helps to manipulate algebraic expressions typically encountered in perturbation theory near Stokes flow. I describe the main idea in \Secref{pattern}, but leave the implementation details and usage to the tutorial notebooks. I walk through a simple example calculation in \Secref{example}.

\section{Algebra by pattern matching}\seclab{pattern}

We consider now solutions to Stokes equations in spherical geometry. As will be clear the method is not limited to spherical geometry, but it is certainly easier to deal with than, say, spheroidal geometry.

In terms of Cartesian tensors, the building block for solutions to Laplace's, and therefore Stokes', equations are
\begin{align}
\frac{r_{i_1}r_{i_2}...r_{i_n}}{r^m}\,.
\end{align}
For example, the fundamental solution to Stokes' equation, the Stokeslet, is
\begin{align}
  G_{i_1i_2} = \frac{\delta_{i_1i_2}}{r} + \frac{r_{i_1}r_{i_2}}{r^3}\,.\eqnlab{stokesletdef}
\end{align}
Suppose we want to compute the derivative of the Stokeslet, $\partial_{i_3}G_{i_1i_2}$:
\begin{align}
  \partial_{i_3}G_{i_1i_2} = \partial_{i_3}\bigg[\frac{\delta_{i_1i_2}}{r} + \frac{r_{i_1}r_{i_2}}{r^3}\bigg]\,.
\end{align}
A trained human immediately recognizes the \emph{pattern} of a derivative of a sum, and knows that the linear operation applies term-wise:
\begin{align}
  = \partial_{i_3}\frac{\delta_{i_1i_2}}{r} + \partial_{i_3}\frac{r_{i_1}r_{i_2}}{r^3}\,.
\end{align}
Focusing on the second term, a human recognizes the \emph{pattern} of a derivative of a product, and applies the product rule:
\begin{align}
  \partial_{i_3}\frac{r_{i_1}r_{i_2}}{r^3}&=
  (\partial_{i_3}r_{i_1})\frac{r_{i_2}}{r^3}
  +r_{i_1}(\partial_{i_3}\frac{r_{i_2}}{r^3})\,,~\textrm{and once more,}\nn\\
&=(\partial_{i_3}r_{i_1})\frac{r_{i_2}}{r^3}
  +r_{i_1}(\partial_{i_3}r_{i_2})\frac{1}{r^3}
  +r_{i_1}r_{i_2}(\partial_{i_3}\frac{1}{r^3})\,.
\end{align}
A human performs these steps intuitively by training and experience, recognizing patterns and replacing them with the appropriate expression according to the rules of algebra and calculus.
The next pattern may be that the derivative $\partial_{j}r_i=\delta_{ij}$, and then perhaps $\delta_{ij}r_j=r_i$.

The idea is that we humans perform calculations by matching patterns in the mathematical expressions to a catalogue of known, legal, rules of algebra and calculus. The goal is to implement a computer program that contains a number of patterns and their associated replacement rules, and let the computer iterate the pattern matching on a mathematical expression until it ceases to change.

In practice, however, we must carefully choose which patterns we implement.
First of all, legal algebraic manipulations such as the examples above, are equalities and therefore valid in both directions. If we were to implement both the pattern $\delta_{ij}r_j\to r_i$, and $r_i \to \delta_{ij}r_j$ the program will get stuck in an infinite loop switching back and forth between the two forms. A human practitioner has an opinion on what form is appropriate, and that opinion must be implemented in the choice of patterns. 
Second, a trained human typically knows a vast catalogue of different patterns and rules applicable in different situations, and many times disambiguate meaning based on context (for example, the implied summation in index notation may be over 2 or 3 or 7 dimensions, known only by context). It is far out of scope to implement any generally valid algebra software, but within a limited context with a reasonable set of mathematical objects, we can make substantial progress.

In the associated Mathematica notebooks I show that by careful choice of patterns and conventions we can implement a program that is useful for calculations in theoretical microhydrodynamics. We already discussed the role of patterns and rules for tensor contraction, differentiation, and so forth. Additionally the program must handle basic symmetry operations. For example, for a symmetric strain tensor $E_{ij}$ we must have $E_{ij}-E_{ji}=0$. Moreover, we must recognize that $E_{ij}r_j-E_{ik}r_k=0$, because dummy indices may be renamed. Finally, the code contains routines to collect terms and prettyprint them in a human-friendly format.

The resulting program is only around 350 lines of code, and the implementation is documented through a sequence of follow-along tutorial notebooks. I refer the interested reader to the tutorial in order to learn the details about syntax and usage.

\section{Example}\seclab{example}

\citet{peery1966} perturbatively calculated the flow and pressure fields around a torque-free rigid spherical particle suspended in a linear flow of a second-order fluid. Here we extend that calculation to include the effects of particle translation, as induced by for example an external force \cite{einarsson2017}. Here we focus on the mathematical labor, and I will not dwell on details about the physics and interpretation.
For convenience I include some screenshots of the program input and output here, but must refer the reader to the accompanying Mathematica notebook for the full program and output.

Consider a spherical particle moving with velocity $\ve v$ and rotating with angular velocity $\ve \omega$ in a linear flow $\ve u^{\infty}=\ve E\ve r+\ve O \ve r$.
The second-order fluid constitutive equation for the stress is
\begin{align}
  \ve \sigma = -\ve \delta p + 2\ve e + \Wi\, \ve\sigma^E,
\end{align}
where
\begin{align}
    \ve\sigma^E &= -2\ve u\cdot\nabla \ve e + 2 (\ve e+\ve o)\ve e + 2\ve e (\ve e-\ve o) + 4\alpha \ve e\ve e \,,\nn\\
  \ve e &= \big(\nabla \ve u + (\nabla \ve u)\tr\big)/2\,,\qquad\ve o = \big(\nabla \ve u - (\nabla \ve u)\tr\big)/2\,.
\end{align}
The momentum balance $\nabla \cdot \ve \sigma=0$ is thus
\begin{align}
  \nabla^2 \ve u - \nabla p = -\Wi \nabla\cdot\ve\sigma^E\,.
\end{align}
The torque-free condition is
\begin{align}
  \int_{S_p} \ve r \times \ve \sigma \ve n \,\rd S=0\,.
\end{align}
The boundary conditions read
\begin{align}
  \ve u &= \ve v + \ve \omega\times \ve r\,\quad\textrm{on the particle surface,}\nn\\
  \ve u &\to \ve u^{\infty}\,\quad |\ve r| \to \infty\,.
\end{align}
At $O(1)$, with $\Wi=0$, this is a textbook Stokes flow problem. We solve it by an ansatz of the Stokeslet and its derivatives:
\begin{align}
 u_i^{(0)} &= u_i^{\infty}+(a_1 + a_2 \nabla^2)G_{ij}v_j + (a_3+\nabla^2 a_4)\partial_kG_{ij}O_{jk}+(a_5+\nabla^2 a_6)\partial_kG_{ij}E_{jk}\,.
\end{align}
This is an implementation of the ansatz in code:
\begin{inlinebox}
\includegraphics[scale=0.5]{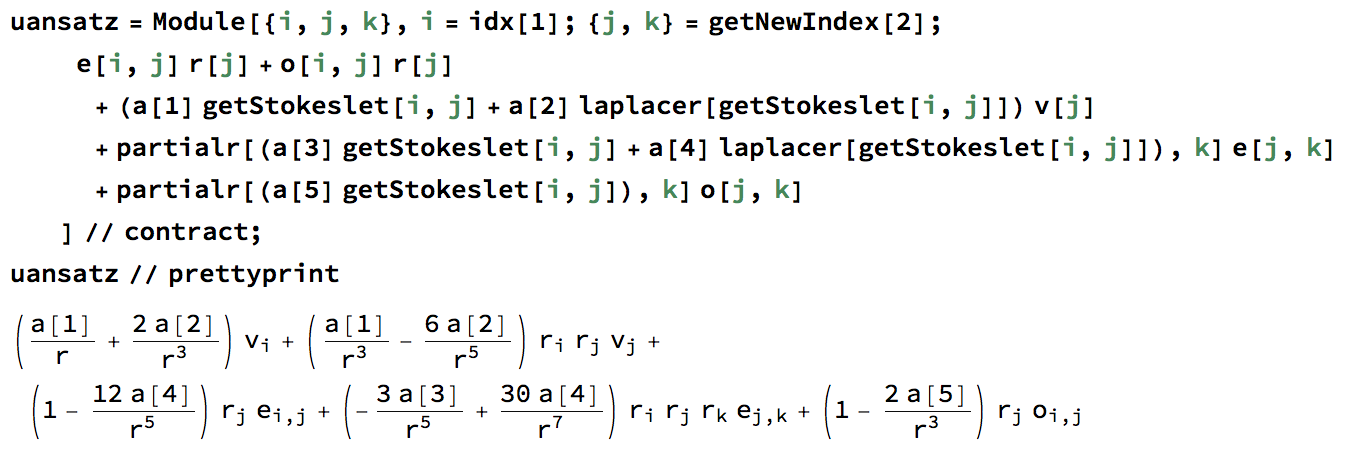}  
\end{inlinebox}
The functions \texttt{partialr[g, i]} and \texttt{laplacer[g]} represent the differential operators $\partial g/\partial r_i$ and $\nabla^2 g$, respectively. The function \texttt{getStokeslet} is simply defined as
\begin{inlinebox}
\includegraphics[scale=0.5]{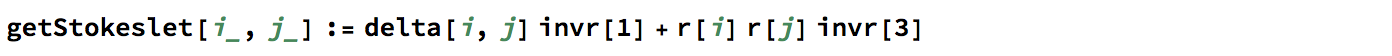} 
\end{inlinebox}
The code is obviously not as dense as mathematical notation, but it is reasonably expressive.
We form the boundary condition
\begin{align}
  \big[u_i^{(0)} - (v_i + O_{ij}r_j) \big]_{r=1}=0\,, \eqnlab{bc0}
\end{align}
where we used that the particle rotates with the flow vorticity to lowest order%
\footnote{This is a textbook result that may elegantly be shown by a reciprocal theorem, or by direct calculation (but that requires the pressure, too).}%
, that is $\ve\omega^{(0)}\times \ve r=\ve O\ve r$. In code this tensorial boundary condition is:
\begin{inlinebox}
\includegraphics[scale=0.5]{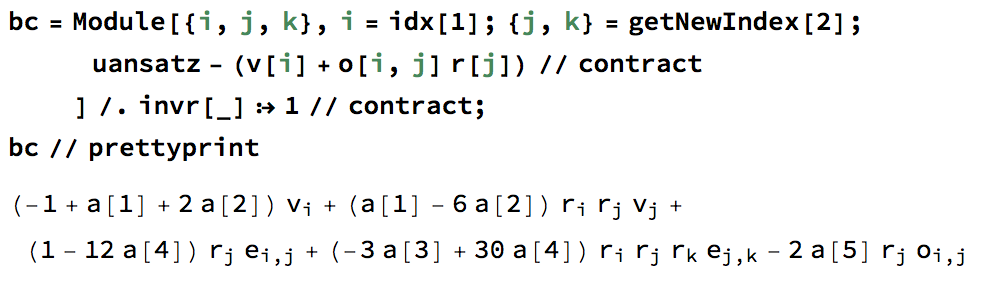}
\end{inlinebox}
The program can collect all scalar prefactors of the unique tensorial groups. All these prefactors must vanish separately, which determines $a_1..a_5$, like so:
\begin{inlinebox}
\includegraphics[scale=0.5]{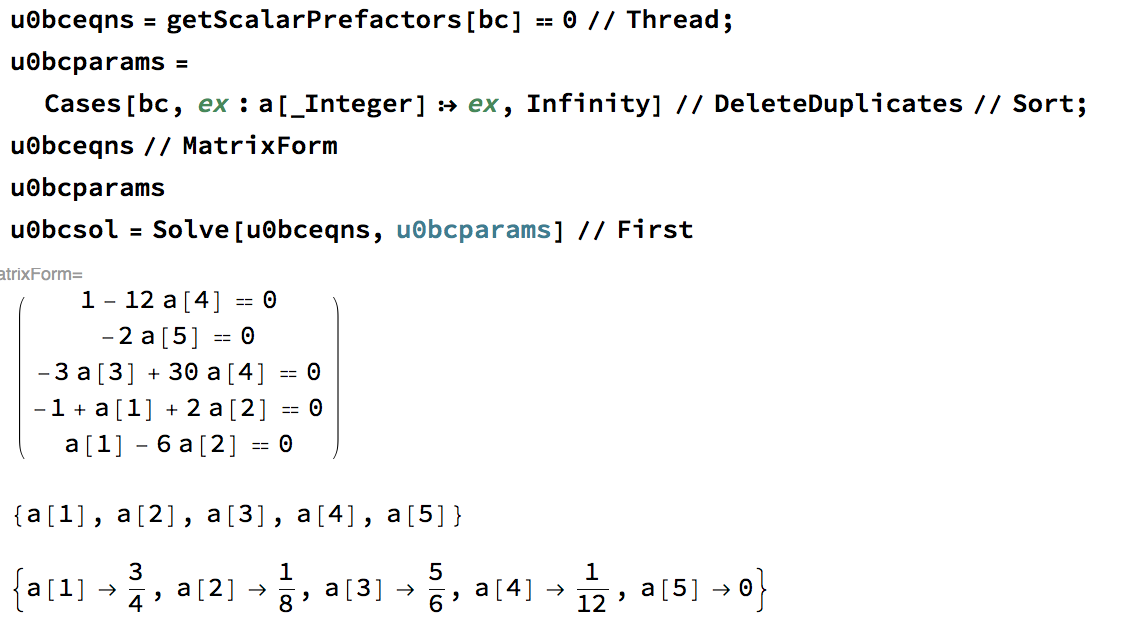}
\end{inlinebox}
Now we may form $\ve u^{(0)}$ by substituting the coefficients:
\begin{inlinebox}
\includegraphics[scale=0.5]{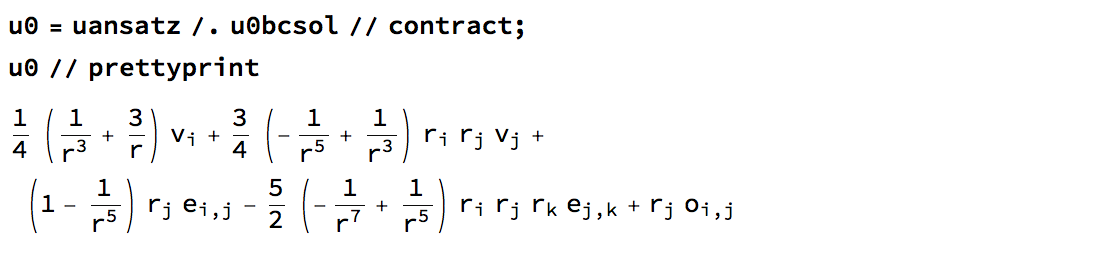}
\end{inlinebox}
Given $u_i^{(0)}$ we can compute $\partial_j u_i^{(0)}$ and thus $e_{ij}^{(0)}$ and $o_{ij}^{(0)}$, the second-order fluid stress, and ultimately the right hand side of the next order equation. The momentum balance at this order is
\begin{align}
  \nabla^2 \ve u^{(1)} - \nabla p^{(1)} = -\nabla\cdot\ve\sigma^{E(0)}\,,\eqnlab{eqn1}
\end{align}
where
\begin{align}
    \ve\sigma^{E(0)} = -2\ve u^{(0)}\cdot\nabla \ve e^{(0)} + 2 (\ve e^{(0)}+\ve o^{(0)})\ve e^{(0)} + 2\ve e^{(0)} (\ve e^{(0)}-\ve o^{(0)}) + 4\alpha \ve e^{(0)}\ve e^{(0)} \,.
\end{align}
\begin{align}
  \ve u^{(1)}&=0\,,\quad\textrm{on the particle surface},\nn\\
  \ve u^{(1)}&\to0\,,\quad |\ve r|\to\infty\,,\eqnlab{bc1}
\end{align}
where we used that the $O(\Wi)$ angular velocity is zero (this can be shown using a reciprocal theorem \cite{einarsson2017}, details omitted here for brevity).

Computing the divergence of the second-order fluid stress is certainly a tedious task, as evident by the pages upon pages of algebra presented by \citet{peery1966}. In the current program, we have
\begin{inlinebox}
\includegraphics[scale=0.5]{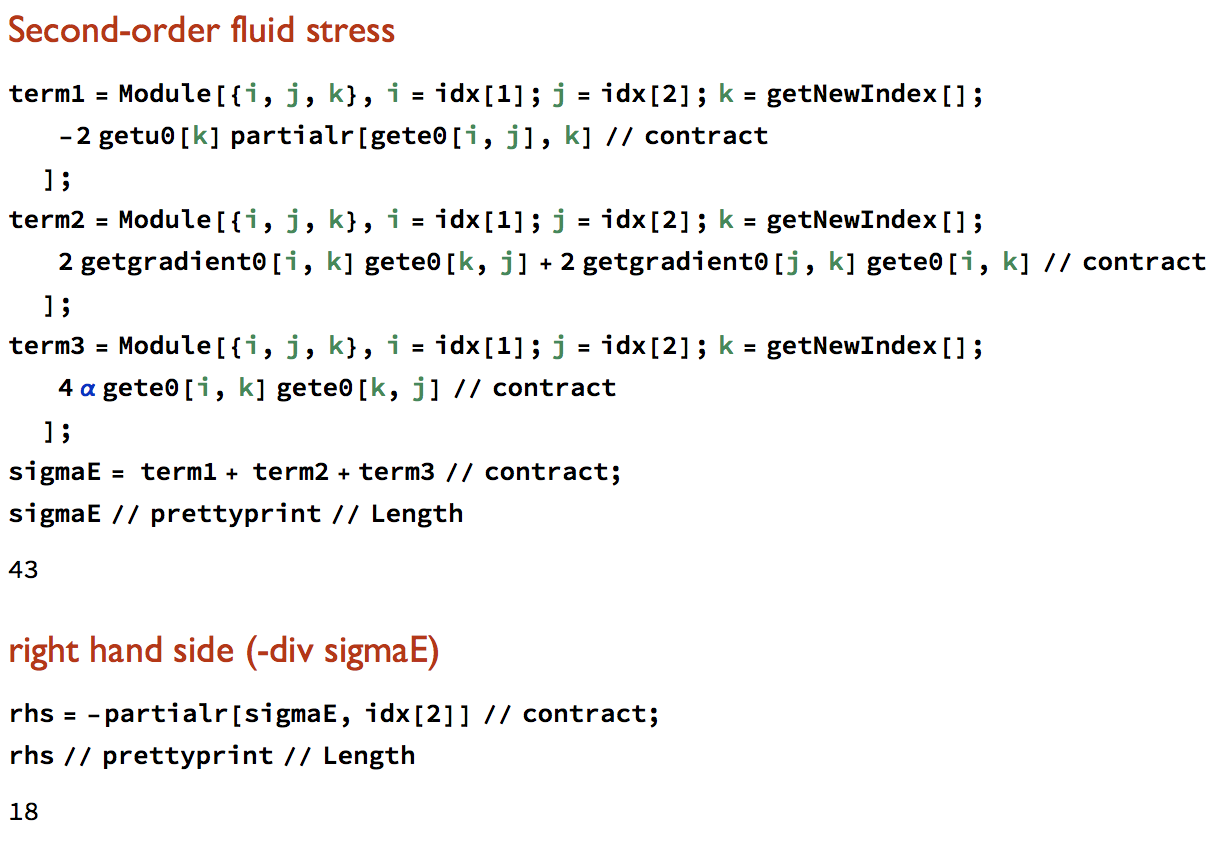}
\end{inlinebox}
That's it. With the inhomogenous Stokes equation \eqnref{eqn1} fully specified, it remains to find a solution. We could proceed like \citet{peery1966}: take the divergence of \Eqnref{eqn1} to form a Poisson equation for the pressure, and solve that by an ansatz of scalar functions $1/r^m$ multiplied by the tensorial forms present in the right hand side of the equation. Then insert the pressure into \Eqnref{eqn1}, and solve the resulting Poisson equation for $\ve u^{(1)}$ with a similar ansatz, such that the boundary conditions are satisfied.

But here, instead, we directly compute a particular solution of \Eqnref{eqn1} by Fourier transform, as shown in Ref.~\citenum{einarsson2017}. In short, a particular solution to
\begin{align}
  \nabla^2 \ve u^{(1)} - \nabla p^{(1)} &= \ve f
\end{align}
is given by
\begin{align}
  \ve u^{(1)p} &= -\invfourier \frac{1}{k^2}(\ve \delta - \frac{\ve k\ve k}{k^2}) \fourier \ve f\,,\eqnlab{u1p}
\end{align}
where $\fourier$ denotes Fourier transform and $\ve k$ is the reciprocal variable. 

The patterns and rules for the Fourier transform of polyadic expressions like $r_ir_j../r^m$ are in fact rather involved, because it turns out that the angular eigenfunctions of the 3D Fourier transform are the spherical harmonics of degree $l$, but the polyadic expressions are typically linear combinations of different values of $l$. Therefore we must separate the parts of the polyadics of different degree $l$, transform, and then recombine. See \citet{einarsson2017} for a detailed derivation. However, once that is implemented, \Eqnref{u1p} is a convenient route to a particular solution, namely:
\begin{inlinebox}
  \includegraphics[scale=0.5]{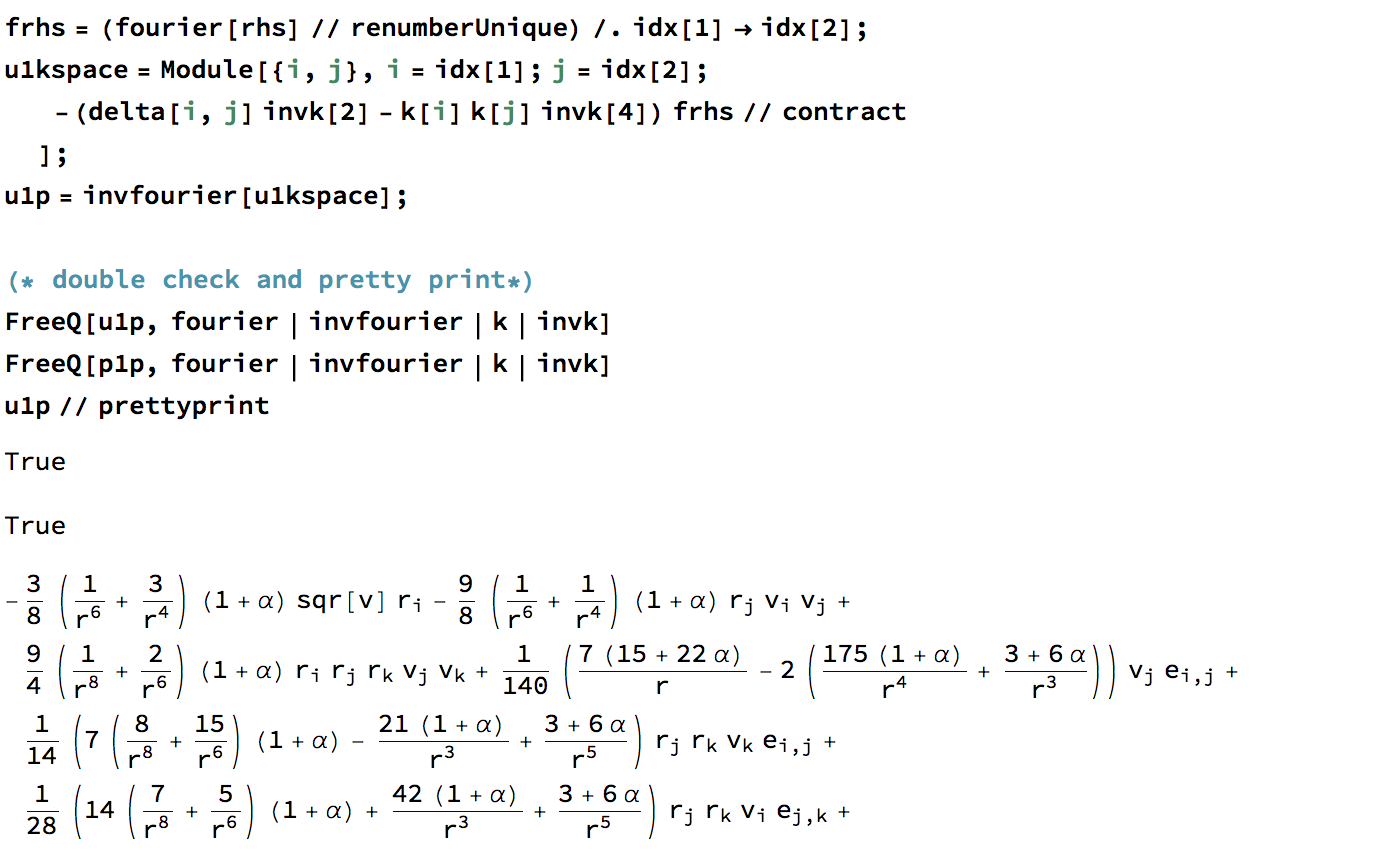}
\end{inlinebox}
\emph{(note that output is cut off in screenshot)}\\
Finally, we add a homogenous solution to $\ve u^{(1)p}$ such that the sum satisfies the boundary conditions \eqnref{bc1}. In this case it is again convenient to make an ansatz of $G_{ij}$ and its derivatives:
\begin{align}
 u_i^{(1)} &= u_i^{(1)p}\nn\\
&+(a_1 + a_2 \nabla^2)\partial_k\partial_l\partial_mG_{ij}E_{jk}E_{lm}\nn\\
&+(a_3 + a_4 \nabla^2)\partial_kG_{ij}E_{jl}E_{lk}\nn\\
&+ \hdots \nn\\
&+(a_{11} + a_{12} \nabla^2)G_{ij}E_{jk}v_k\nn\\
&+(a_{13} + a_{14} \nabla^2)G_{ij}O_{jk}v_k\nn\\
&+(a_{15} + a_{16} \nabla^2)\partial_kG_{ij}v_jv_k\,.
\end{align}
See the notebook for the implementation of this ansatz. The program readily simplifies and evaluates this expression at $r=1$:
\begin{inlinebox}
  \includegraphics[scale=0.5]{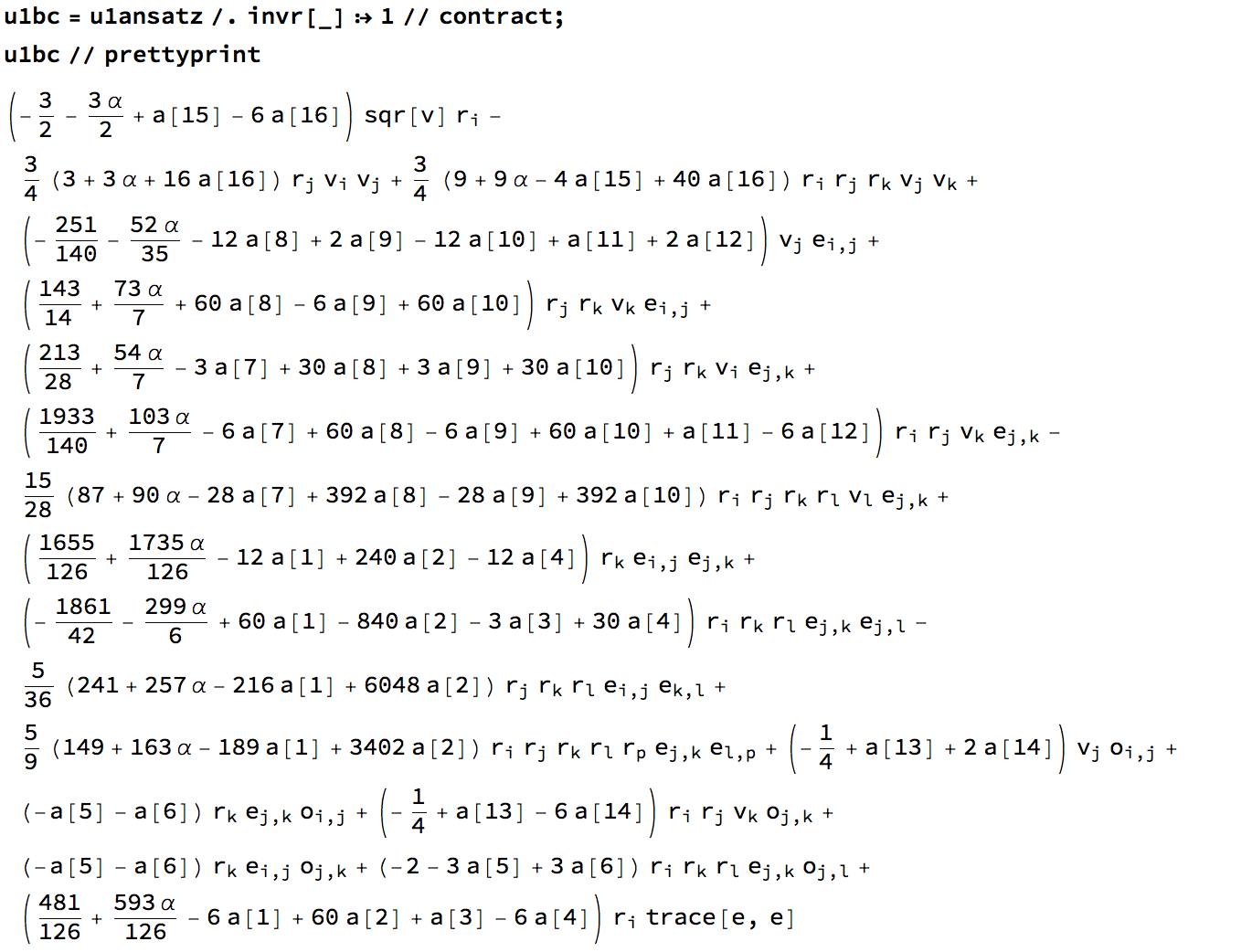}
\end{inlinebox}
Similarly to the lowest order, but more complicated, we require the scalar prefactors of all tensorial expressions to vanish independently:
\begin{inlinebox}
\includegraphics[scale=0.5]{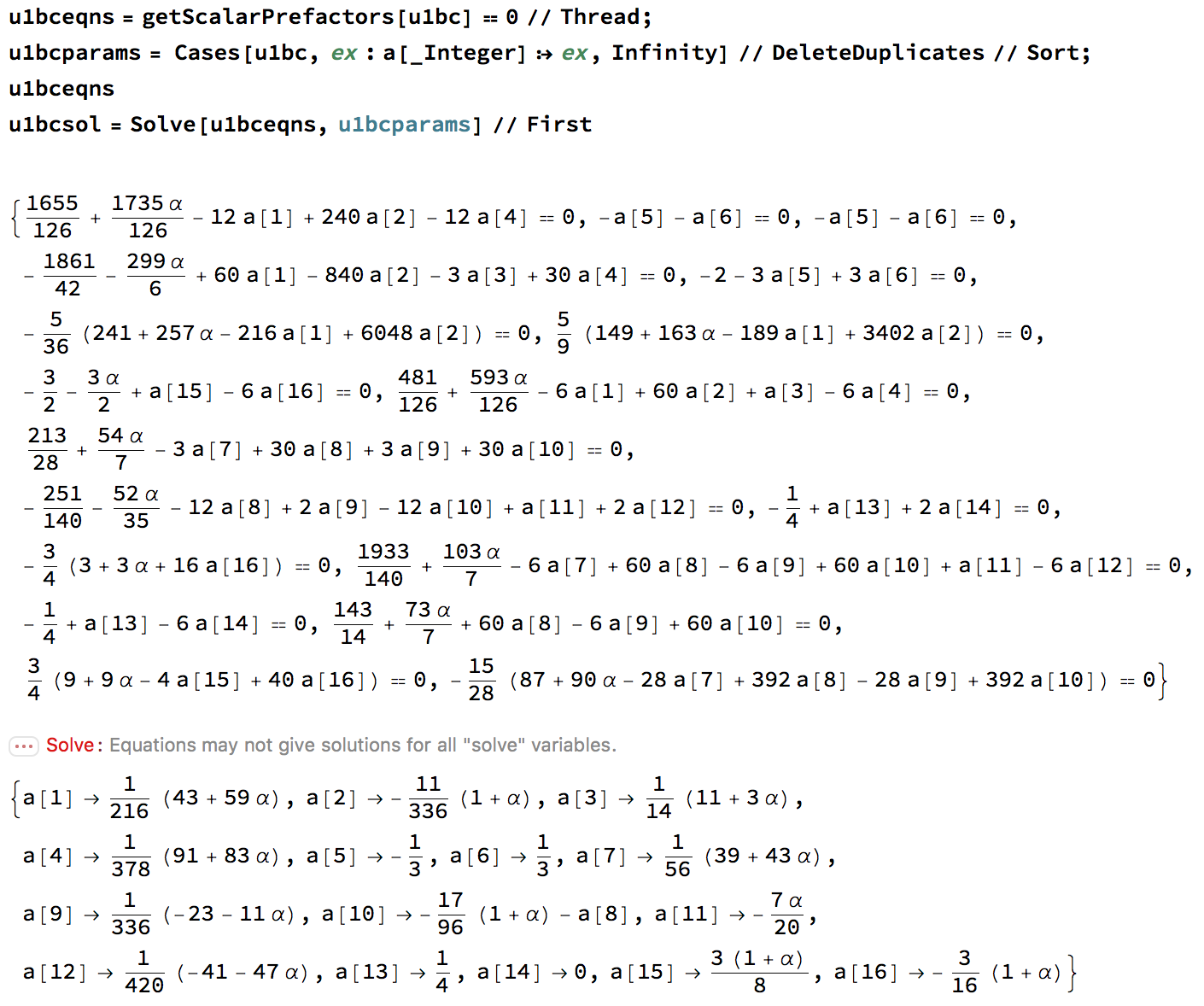}
\end{inlinebox}
This determined $a_1..a_{16}$, and therefore $\ve u^{(1)}$:
\begin{inlinebox}
\includegraphics[scale=0.5]{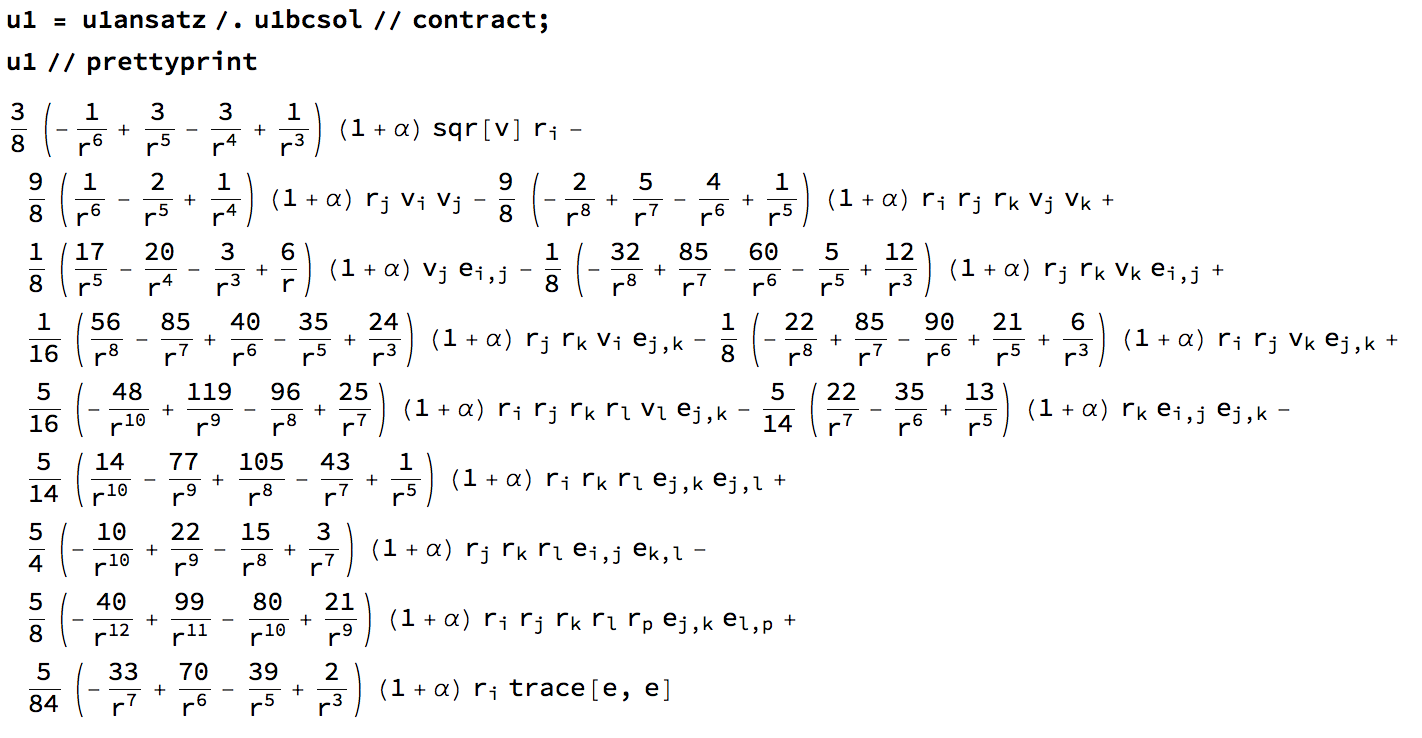}
\end{inlinebox}
In the case of $\ve v = 0$, we recover the result of \citet{peery1966} [Eq.~(4.23)]:
\begin{inlinebox}
\includegraphics[scale=0.5]{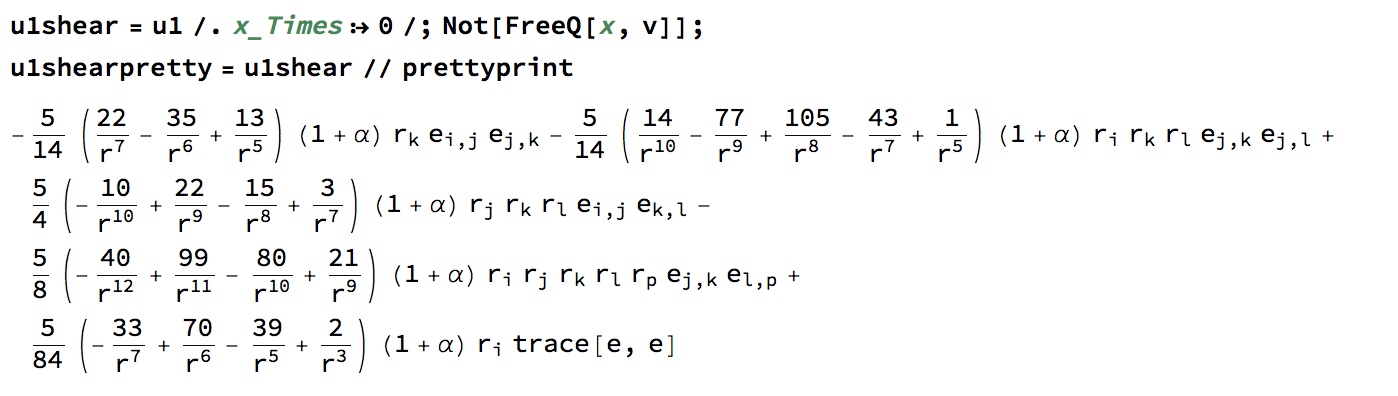}
\end{inlinebox}
The runtime of this entire calculation is about ten seconds on a 2016 laptop.

\section{Discussion}

In this note I have presented a method of computer algebra that helps with some of the most laborious tasks in theoretical microhydrodynamics. Perhaps this demonstration can inspire attempts at new results by computer algebra, and push the limit of how complicated models we tackle analytically.

The examples shown in this note all pertain to microhydrodynamics in spherical geometry, but it should be clear that the general method applies as long as one can invent patterns and rules for the mathematical objects in question. For example, I employed an early version of this method for my thesis work on the rotation rate of spheroidal particles \cite{einarsson2015a}. In that case the solutions to Stokes' equation are polyadics in the position vector $\ve r$, and the spheroid axis of symmetry $\ve n$, multiplied with the scalar functions
\begin{align}
  I_m^n(|\ve r|, \ve r\cdot \ve n)&=\int_{-c}^c \frac{\xi^n}{|\ve r - \xi \ve n|^m}\rd\xi\,,
\end{align}
where $c$ is a constant related to the spheroidal geometry. While these scalar functions are much more complicated than the spherical counterpart $1/r^m$, they still obey certain rules, for example Eq.~(B3) in \cite{einarsson2015a}:
\begin{align}
  \frac{\partial}{\partial n_i}I_m^n&=m r_i I_{m+2}^{n+1} - m n_i I_{m+2}^{n+2}\,.
\end{align}
The point is, if you can perform the algebra by pattern matching and replacement, this method can help with the mechanical labor.

\begin{acknowledgements}
I would like to thank Kristian Gustafsson who once upon a time got me started on the pattern matching path, and Joe Barakat for inspiring me to finally clean up and release this code.
\end{acknowledgements}

\bibliography{refs}
\end{document}